\documentclass[usenatbib,referee]{mn2e}
\usepackage{epsf}
\usepackage{graphicx}

\def\eps@scaling{1.0}%
\newcommand\epsscale[1]{\gdef\eps@scaling{#1}}%
\newcommand\plotone[1]{%
 \centering
 \leavevmode
 \includegraphics[width={\eps@scaling\columnwidth}]{#1}%
}%

\def\apj{ApJ}

\def\beq{\begin{equation}}
\def\eeq{\end{equation}}
\def\bey{\begin{eqnarray}}
\def\eey{\end{eqnarray}}

\def\mpc{\,h^{-1}{\rm {Mpc}}}

\def\kpc{\,h^{-1}{\rm {kpc}}}

\def\Msun{{M_\odot}}

\def\gs{\mathrel{\raise1.16pt\hbox{$>$}\kern-7.0pt
\lower3.06pt\hbox{{$\scriptstyle \sim$}}}}
\def\ls{\mathrel{\raise1.16pt\hbox{$<$}\kern-7.0pt
\lower3.06pt\hbox{{$\scriptstyle \sim$}}}}
\def\gtsima{$\; \buildrel > \over \sim \;$}
\def\ltsima{$\; \buildrel < \over \sim \;$}
\def\prosima{$\; \buildrel \propto \over \sim \;$}
\def\gsim{\lower.5ex\hbox{\gtsima}}
\def\lsim{\lower.5ex\hbox{\ltsima}}
\def\simgt{\lower.5ex\hbox{\gtsima}}
\def\simlt{\lower.5ex\hbox{\ltsima}}
\def\simpr{\lower.5ex\hbox{\prosima}}
\def\la{\lsim}
\def\ga{\gsim}

\title
[Environmental Dependence of Cold Dark Matter Halo Formation]
{Environmental Dependence of Cold Dark Matter Halo Formation}
\author[H. Wang et al.]
   {\parbox[t]{\textwidth}{
       H.Y. Wang$^{1,3,4}$\thanks{E-mail: whywang@mail.ustc.edu.cn},
       H.J. Mo$^{2,3}$,
       Y.P. Jing$^{3,4}$
}\\
            $^1$Center for Astrophysics, University of Science and
                Technology of China, Hefei, Anhui, China\\
            $^2$Department of Astronomy, University of Massachusetts,
            Amherst MA 01003-9305, USA\\
            $^3$Shanghai Astronomical Observatory; the Partner Group of MPA,
            Nandan Road 80, Shanghai 200030, China \\
            $^4$Joint Institute for Galaxy and Cosmology (JOINGC) of SHAO
            and USTC}

\date{
Accepted ........
Received .......;
in original form ......}

\pubyear{2006}
\begin{document}
\maketitle \label{firstpage}

\label{firstpage}

\begin{abstract}
We use a high-resolution $N$-body simulation to study how the
formation of cold dark matter (CDM) halos is affected by their
environments, and how such environmental effects produce the
age-dependence of halo clustering observed in recent $N$-body
simulations. We estimate, for each halo selected at redshift
$z=0$, an `initial' mass $M_{\rm i}$ defined to be the mass
enclosed by the largest sphere which contains the initial barycenter of
the halo particles and within which the mean linear density is
equal to the critical value for spherical collapse at $z=0$.
For halos of a given final mass, $M_{\rm h}$, the ratio $M_{\rm
i}/M_{\rm h}$ has large scatter, and the scatter is larger for
halos of lower final masses. Halos that form earlier on average
have larger $M_{\rm i}/M_{\rm h}$, and so correspond to higher
peaks in the initial density field than their final masses imply.
Old halos are more strongly clustered than younger ones of the
same mass because their initial masses are larger. The
age-dependence of clustering for low-mass halos is entirely due to
the difference in the initial/final mass ratio. Low-mass old halos
are almost always located in the vicinity of big structures, and
their old ages are largely due to the fact that their mass
accretions are suppressed by the hot environments produced by the
tidal fields of the larger structure. The age-dependence of
clustering is weaker for more massive halos because the heating by
large-scale tidal fields is less important.
\end{abstract}


\section{Introduction}

In the standard cold  dark matter (CDM) paradigm of structure
formation, virialized CDM  halos are both the building blocks of
the large-scale structure of the Universe and the hosts within
which galaxies are supposed to form. During the last decade, the
properties of CDM halos, such as their internal structure,
formation histories and clustering properties,  have been studied
in great detail using both $N$-body numerical simulations and
analytical models. These studies have shown that the halo bias
depends strongly on halo mass, in the sense that more massive
halos are more strongly clustered (e.g., Mo \& White 1996; Jing
1998; Seljak \& Warren 2004). This mass dependence of  the halo
bias has played a crucial role in understanding  the correlation
function of both dark  matter and  galaxies,  via the halo model
(e.g., Mo, Jing \& B\"orner 1997; Ma \& Fry 2000; Seljak 2000),
the halo occupation models (e.g., Jing, Mo \& B\"{o}rner 1998;
Peacock \& Smith 2000; Zheng et al. 2005), and the conditional
luminosity function (e.g., Yang, Mo \& van den Bosch 2003).

 Recently Gao et al (2005), using a very large $N$-body simulation
of a $\Lambda$CDM cosmology, reexamined  the dependence of halo
bias on the properties of dark matter halos. They found that,  for
low-mass halos at redshift $z=0$ with $M_{\rm h} \ll M_\star$
\footnote{$M_\star$ is the characteristic mass scale at which the
{\it RMS} of the linear density field is equal to $1.686$ at the
present time. For the present simulation $M_\star\approx 1.0\times
10^{13}h^{-1}{\rm M}_\odot$.}, the bias depends not only on the
mass but also on the formation time of dark matter halos. For a
given mass, halos that have earlier formation times are on average
more strongly clustered in space. This age-dependence of halo
clustering has been confirmed by a number of independent studies
(e.g. Zhu et al. 2006; Harker et al. 2006). Since halo formation
time is known to be correlated with halo concentration (e.g.
Navarro, Frenk \& White 1997; Jing 2000; Wechsler et al 2002; Zhao
et al. 2003a,b; Lu et al. 2006), the age dependence of halo
clustering also shows up as a halo-concentration dependence of
halo clustering (Wechsler et al. 2005). For massive halos with
$M_{\rm h}\gg M_\star$, the age-dependence may be reversed, as
shown in Wechsler et al. (2005) and Wetzel et al. (2006).

 Although the mass-dependence of halo clustering is well
understood in the excursion-set model based on spherical collapse
in Gaussian initial density field (Mo \& White 1996), the
formation-time dependence (or age-dependence)
is not. In the simplest excursion-set
model from which the Press-Schechter formula is derived (Bond et
al. 1991), the formation history of a halo is entirely determined
by the local density field, independent of the mass distribution
on larger scales. Therefore, the large-scale clustering of halos
is independent of formation time in this model. An age-dependence
of halo clustering is expected in the excursion-set model based on
ellipsoidal collapse (Sheth, Mo \& Tormen 2001), because here the
collapse of a halo depends not only on the local density field but
also on the tidal field generated by its large-scale environment.
Unfortunately, this aspect of the model was not explored in any
detail in Sheth et al. (2001). Instead, these authors used a mean
relation between halo mass and the shape parameter of the
large-scale tidal field to derive an average critical collapse
overdensity that depends only on halo mass.

  Since the properties of galaxies that form in a halo are
expected to depend on the formation history of the halo,
understanding the origin of the age-dependence of halo clustering
is crucial for a better understanding about how galaxies are
related to their large-scale environments. In this paper we use
a high-resolution $N$-body simulation to study how the formation of
dark matter halos is affected by their environments, and how such
environmental effects can be used to understand the age-dependence
of halo clustering. The outline of the paper is as follows. In
Section \ref{sec:simulation} we describe briefly the simulation to be
used and how dark halos are identified. In Section \ref{sec:age}
we study the age-dependence of halo clustering in terms of halo
properties in the initial density field. In Section \ref{sec:env}
we examine the environments of dark halos and how they may affect
the formation histories of dark matter halos. We discuss and
summarize our results in Section \ref{sec:discussion}.

\section{Simulation and dark matter halos}
\label{sec:simulation}

The simulation used in this paper was obtained using the ${\rm
P^3M}$ code described in Jing \& Suto (2002). This simulation
assumes a spatially-flat $\Lambda$CDM model, with density
parameters $\Omega_m=0.3$ and $\Omega_\Lambda=0.7$, and the cold
dark matter power spectrum given by Bardeen et al (1986), with a
shape parameter $\Gamma=\Omega_mh=0.2$ and an amplitude specified
by $\sigma_8=0.9$. The CDM density field was represented by
$512^3$ particles, each having a mass of $M_p \sim 6.2 \times
10^{8}\Msun$, in a cubic box of 100 $\mpc$. The softening length
is $10 \kpc$. Outputs are made at 60 different redshifts between
$z=15$ and $z=0$, with an interval given by $\Delta\ln (1+z)=0.047$.

\begin{figure*}
\plotone{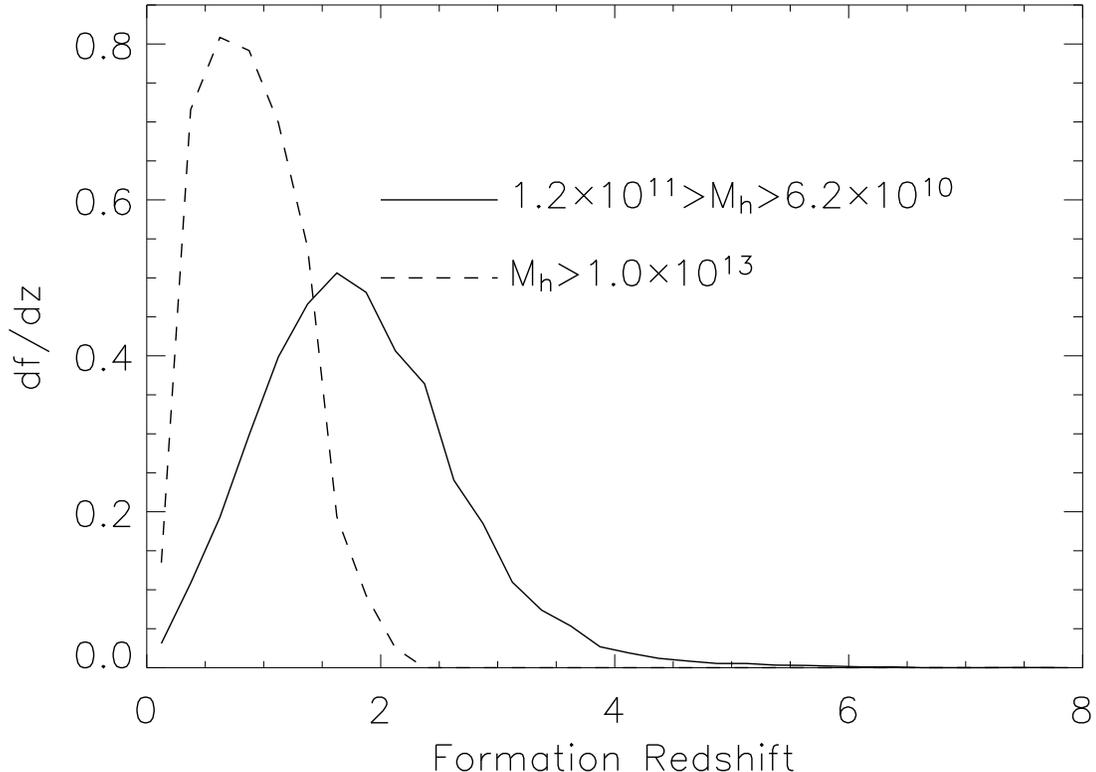} \caption{The distribution of formation redshift
for halos in two mass ranges, $1.2\times 10^{11} h^{-1}{\rm
M}_\odot>M_h>6.2\times 10^{10} h^{-1}{\rm M}_\odot$ (solid curve)
and $M_h> 1.0 \times 10^{13}h^{-1}{\rm M}_\odot$ (dashed curve).}
\label{fig:age_distribution}
\end{figure*}

Halos are identified with the friends-of-friends algorithm (e.g.
Davis et al. 1985) with a linking length $0.2$ times the mean
particle separation. In order to ensure that only physically bound
systems are selected, we discard all particles selected into a
halo but are not bound on the basis of their binding energies.
However, this step does not have any significant impact on our results.
Halos at $z=0$ are related to their progenitors at higher $z$
through halo merging trees. A halo in an earlier output is
considered to be a progenitor of the halo in question if more than
half of its particles are found in the final halo. The formation
time of a halo is defined as the time when its most massive
progenitor has exactly half of the final mass. Interpolations
between adjacent outputs were adopted when estimating the
formation times. In Fig.\,\ref{fig:age_distribution} we show the
distribution of formation redshift for halos in two mass ranges.
These results are similar to those obtained in Li et al. (2005).

\section{The age dependence of halo clustering}
\label{sec:age}

\begin{figure*}
\plotone{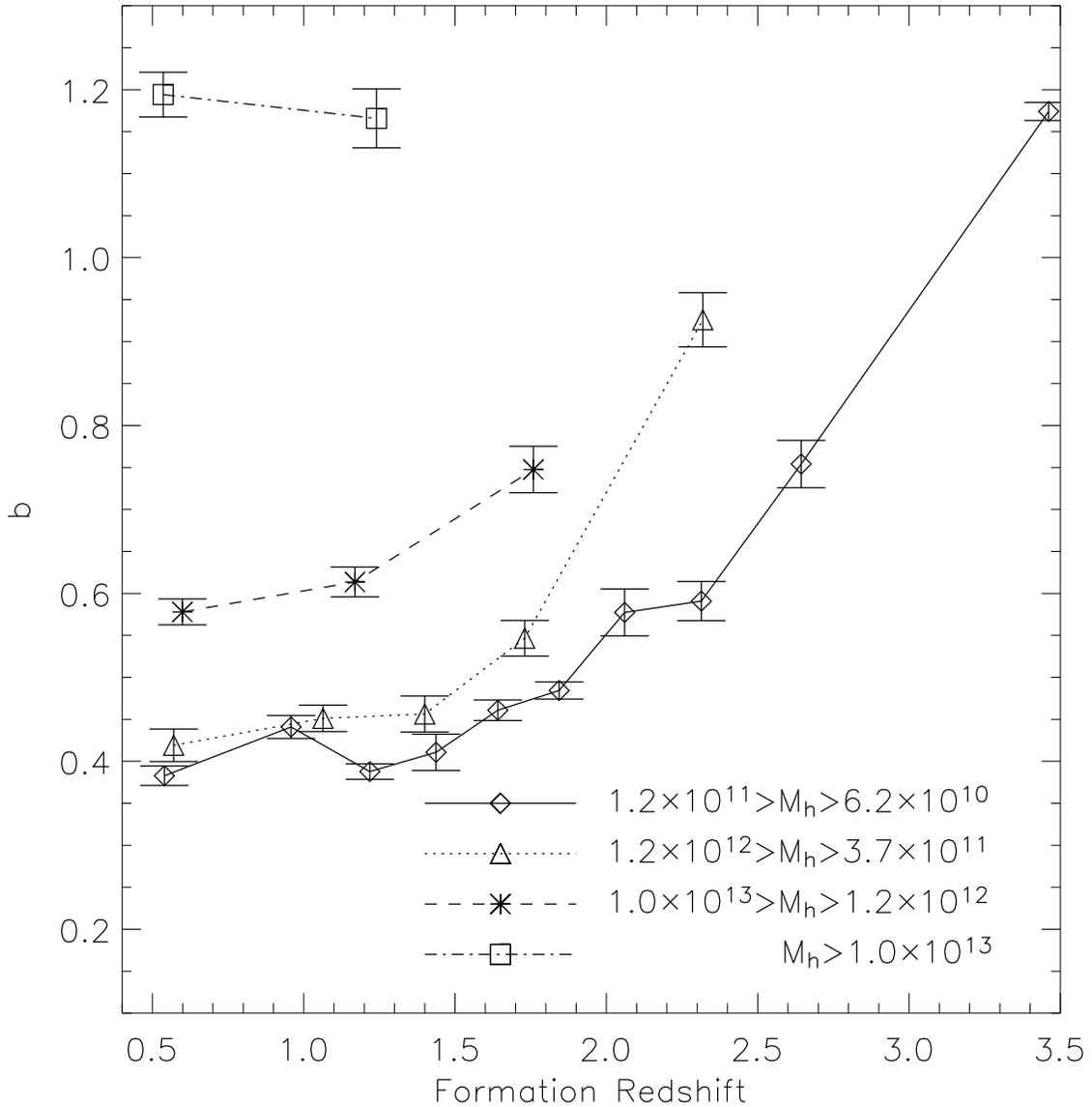} \caption{The bias factor, defined by Eq.
\ref{eq:bias}, as a function of halo formation redshift. Results
are shown for halos in different mass ranges, as indicated in the
panel.} \label{fig:1}
\end{figure*}

  In order to examine the age-dependence of halo clustering
in the present simulation, we estimate the mean overdensity of
dark matter within a sphere of a given radius $R$ around each dark
matter halo, $\delta_h (R)\equiv M(R)/{\overline M}(R) -1$, where
$M(R)$ is the mass within a radius $R$ around the halo, and
${\overline M}(R)=(4\pi R^3/3){\overline \rho}$ with
${\overline\rho}$ being the mean mass density of the universe. We
estimate the bias parameter of a given set of halos using
\begin{equation}\label{b_definition}
b={\langle\delta_h (R)\rangle\over \langle\delta_m
(R)\rangle}\,,\label{eq:bias}
\end{equation}
where $\langle\delta_h (R)\rangle$ is the average overdensity over
the set of halos in consideration, and $\langle\delta_m
(R)\rangle$ is the average overdensity within all spheres of
radius $R$ centered on dark matter particles. This
definition of $b$ is different from that in Gao et al. (2005)
where the auto-correlation function of halos is used to
derive the bias factor. Our test showed that these two
definitions lead to similar results. The advantage of
our definition is that the statistic is more stable
owing to the fact that there are many more dark matter
particles than halos in the simulation. Fig.\,\ref{fig:1}
shows the bias parameter obtained in this way as a function of
halo formation time for halos of different masses. Results shown
are the average for five different choices of $R$: $5$, $7.5$,
$10$, $12.5$ and $15\,h^{-1}{\rm Mpc}$. We see clearly that the
bias factor increases with halo formation redshift for low-mass
halos ($M_h< M_\star$) and the dependence becomes weaker for
$M_h\sim M_\star$. These results are in good agreement with those
obtained by Zhu et al. (2006), using the same set of simulations,
from the auto-correlation functions of dark halos. For massive
halos, there is indication that the age-dependence may be
reversed. Unfortunately, the present simulation is too small to
allow us to study the age-dependence in detail for massive halos
with $M_h > 10^{13} h^{-1}{\rm M}_\odot$. Note that part of the
inverse relation between halo bias and formation time for massive
halos is due to the use of non-zero mass bin. More massive halos
within such a bin have, on average, larger bias due the the strong
mass-dependence of halo bias for massive halos. Since the more
massive halos are on average also younger, an inverse relation
between halo bias and formation time can be produced, even if
age-dependence of halo bias is completely absent for massive halos
of a given mass.

\begin{figure*}
\epsscale{0.4}\plotone{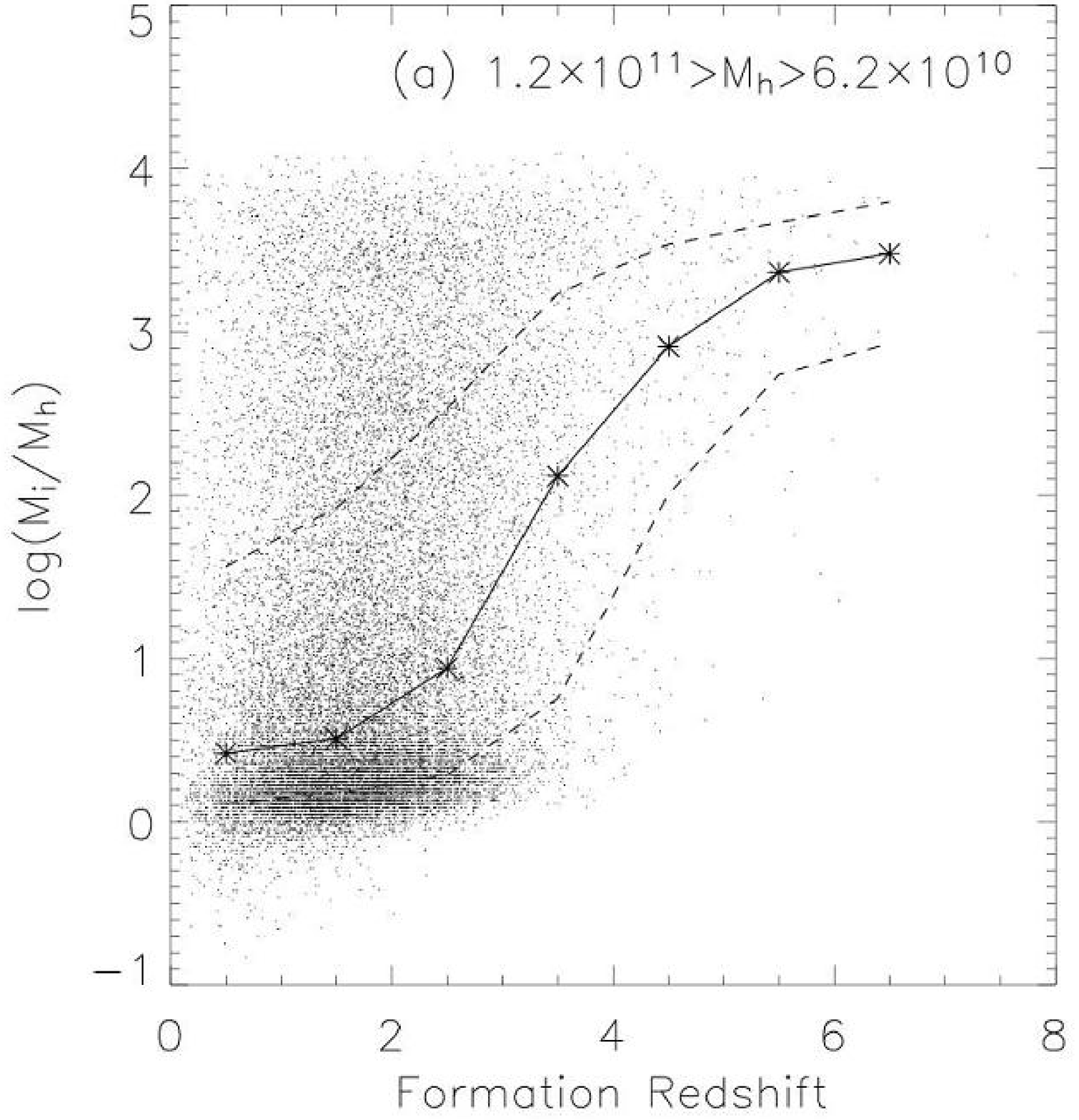}\plotone{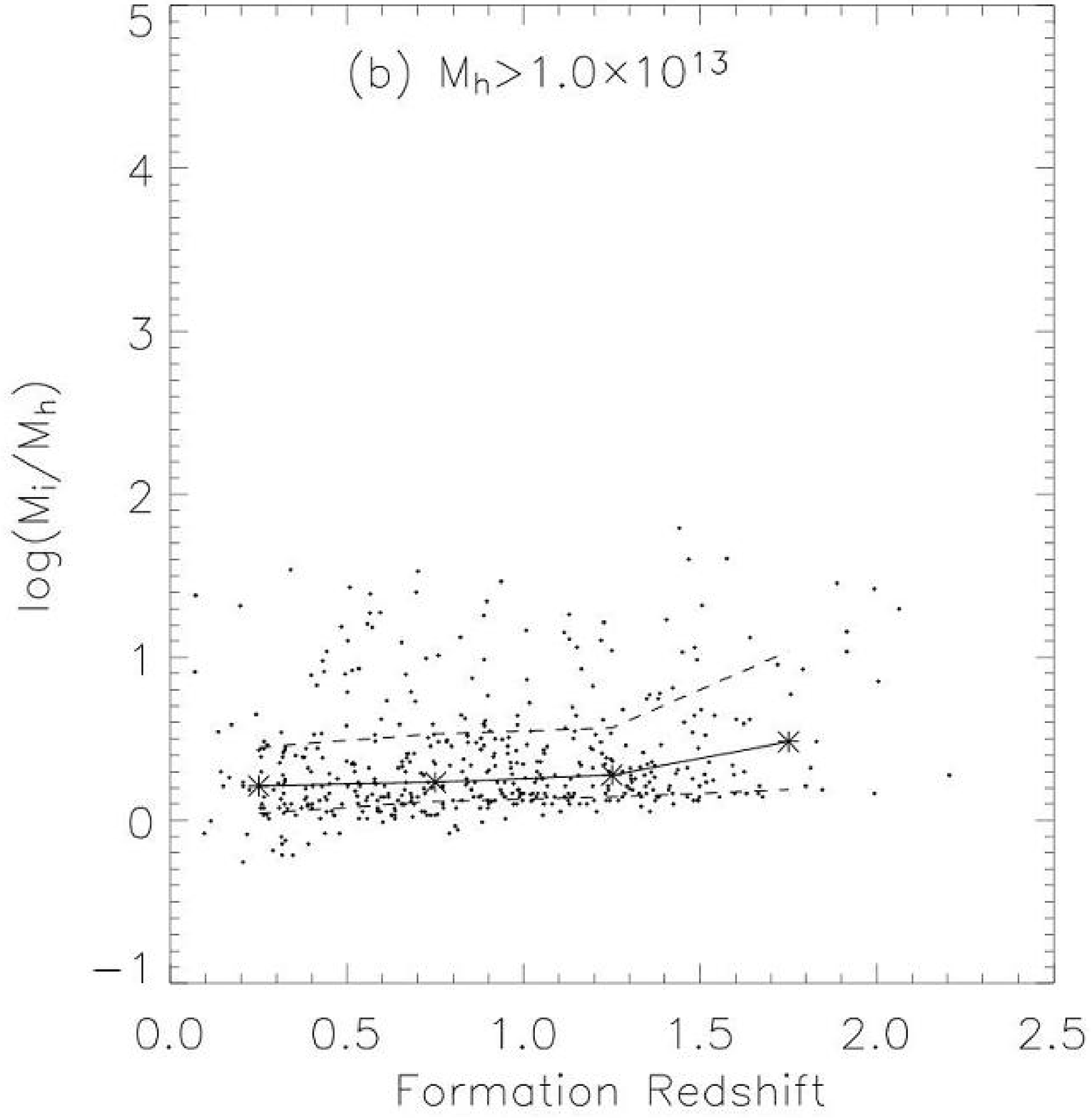} \caption{The
ratio between initial mass, $M_{\rm i}$, and the final mass $M_h$
as a function of halo formation redshift. Left panel is for
low-mass halos with $1.2\times 10^{11} h^{-1}{\rm
M}_\odot>M_h>6.2\times 10^{10} h^{-1}{\rm M}_\odot$, while right
panel is for halos with $M_h> 1.0 \times 10^{13}h^{-1}{\rm
M}_\odot$. The three curves in each panel represent the median, 20
and 80 percentiles.} \label{fig:2}
\end{figure*}
\begin{figure*}
\epsscale{1}\plotone{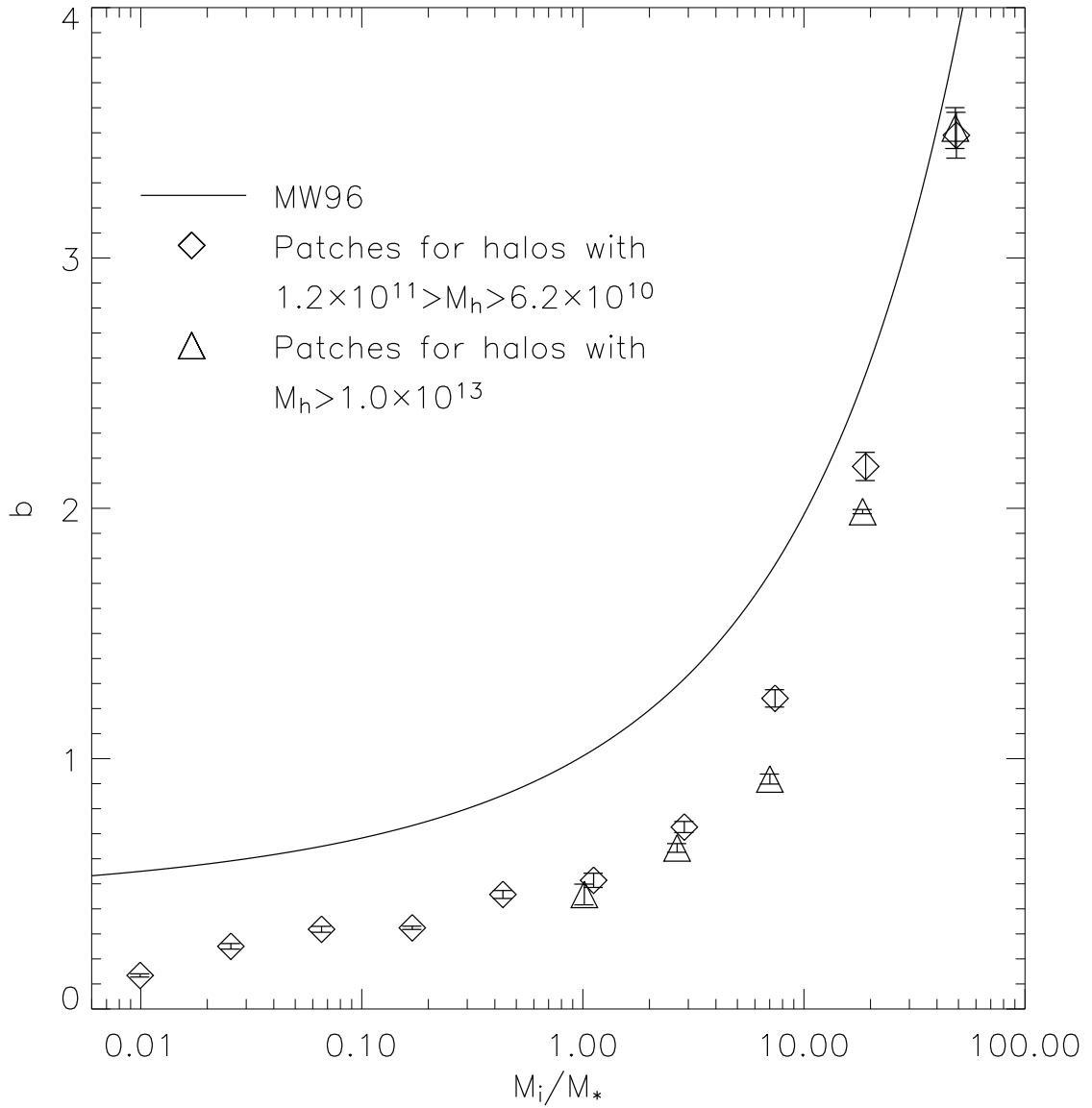} \caption{Bias factor as a function of
initial mass $M_{\rm i}$ for low-mass halos with $1.2\times
10^{11} h^{-1}{\rm M}_\odot>M_h>6.2\times 10^{10} h^{-1}{\rm
M}_\odot$ (squares), and massive halos with $M_h> 1.0 \times
10^{13}h^{-1}{\rm M}_\odot$ (triangles). The solid curve shows the
prediction of the Mo \& White (1996) bias model.} \label{fig:3}
\end{figure*}
\begin{figure*}
\plotone{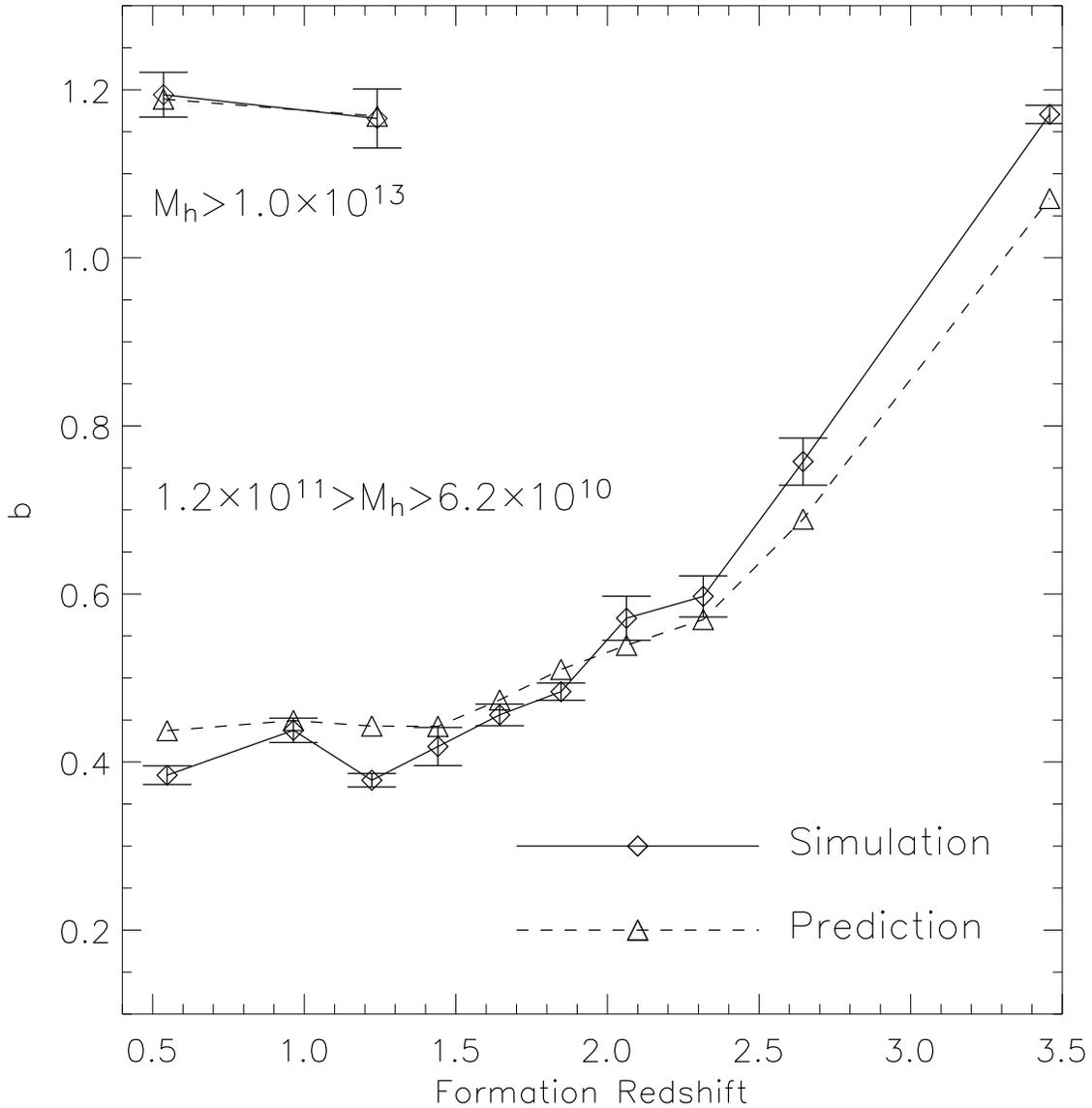} \caption{Bias factor as a function of formation
redshift. Symbols connected with solid lines are the results
obtained from simulations (the same as those shown in
Fig.\,\ref{fig:1}), while symbols connected by dashed lines are
the corresponding predictions based on the correlation between
bias and initial mass (see text for detail). Results are shown for
low-mass halos with $1.2\times 10^{11} h^{-1}{\rm
M}_\odot>M_h>6.2\times 10^{10} h^{-1}{\rm M}_\odot$, and for
massive halos with $M_h> 1.0 \times 10^{13}h^{-1}{\rm M}_\odot$,
as indicated in the panel. For the other two mass ranges shown in Fig 2,
the model predictions match the simulation results as well.
They are not plotted here for the sake of clarity.}
\label{fig:4}
\end{figure*}
\begin{figure*}
\epsscale{0.4}\plotone{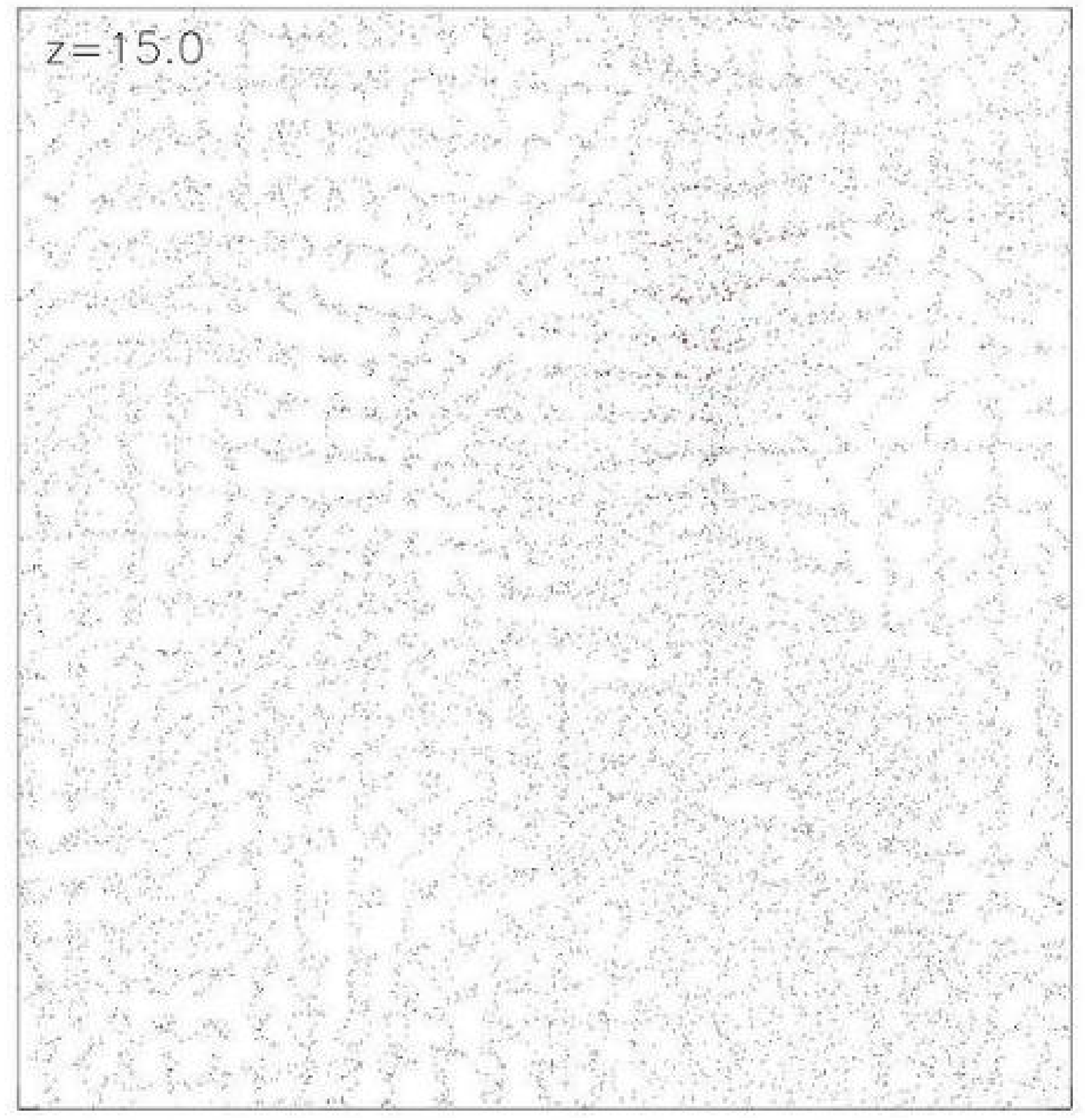}\plotone{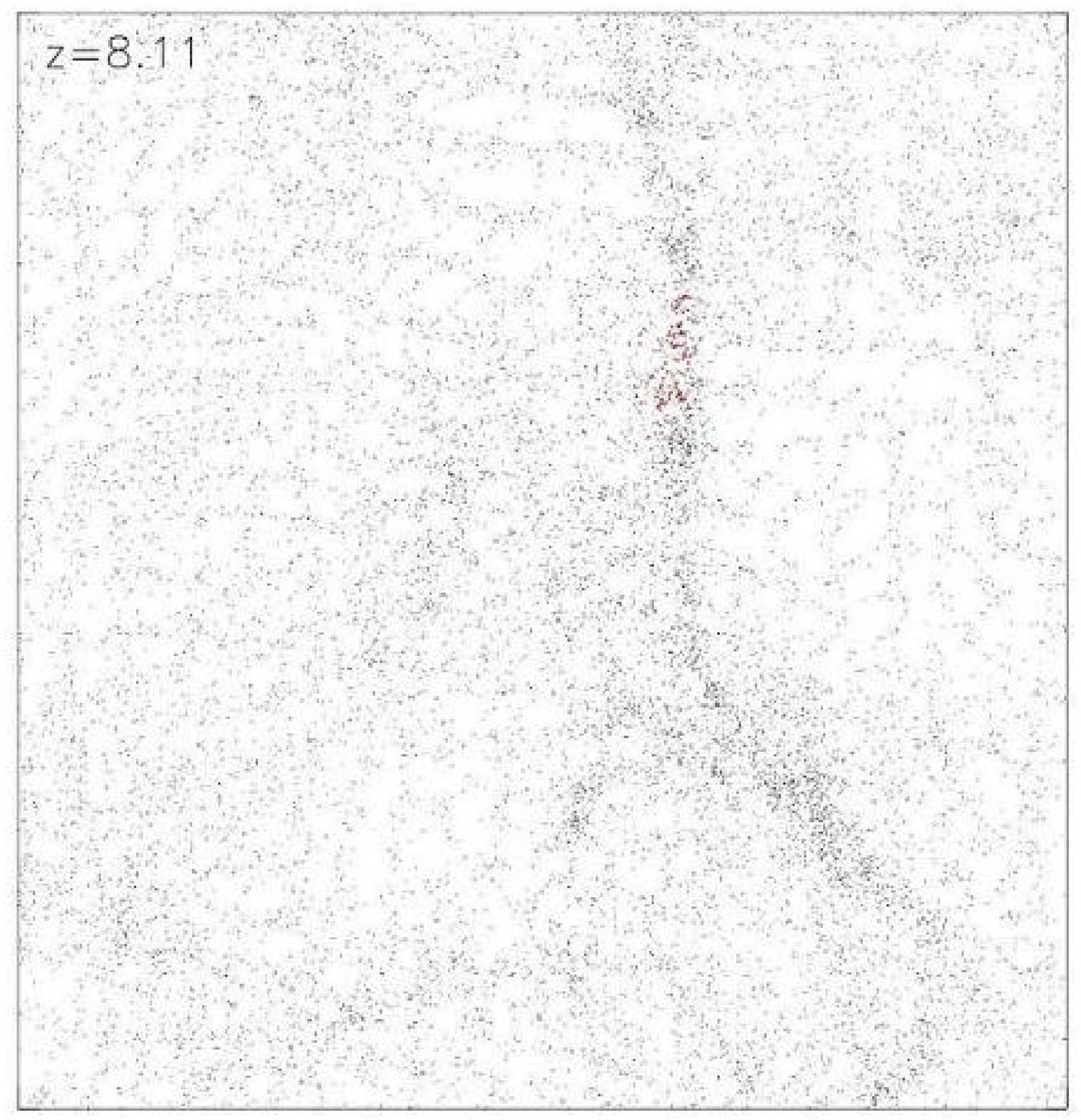}
\epsscale{0.4}\plotone{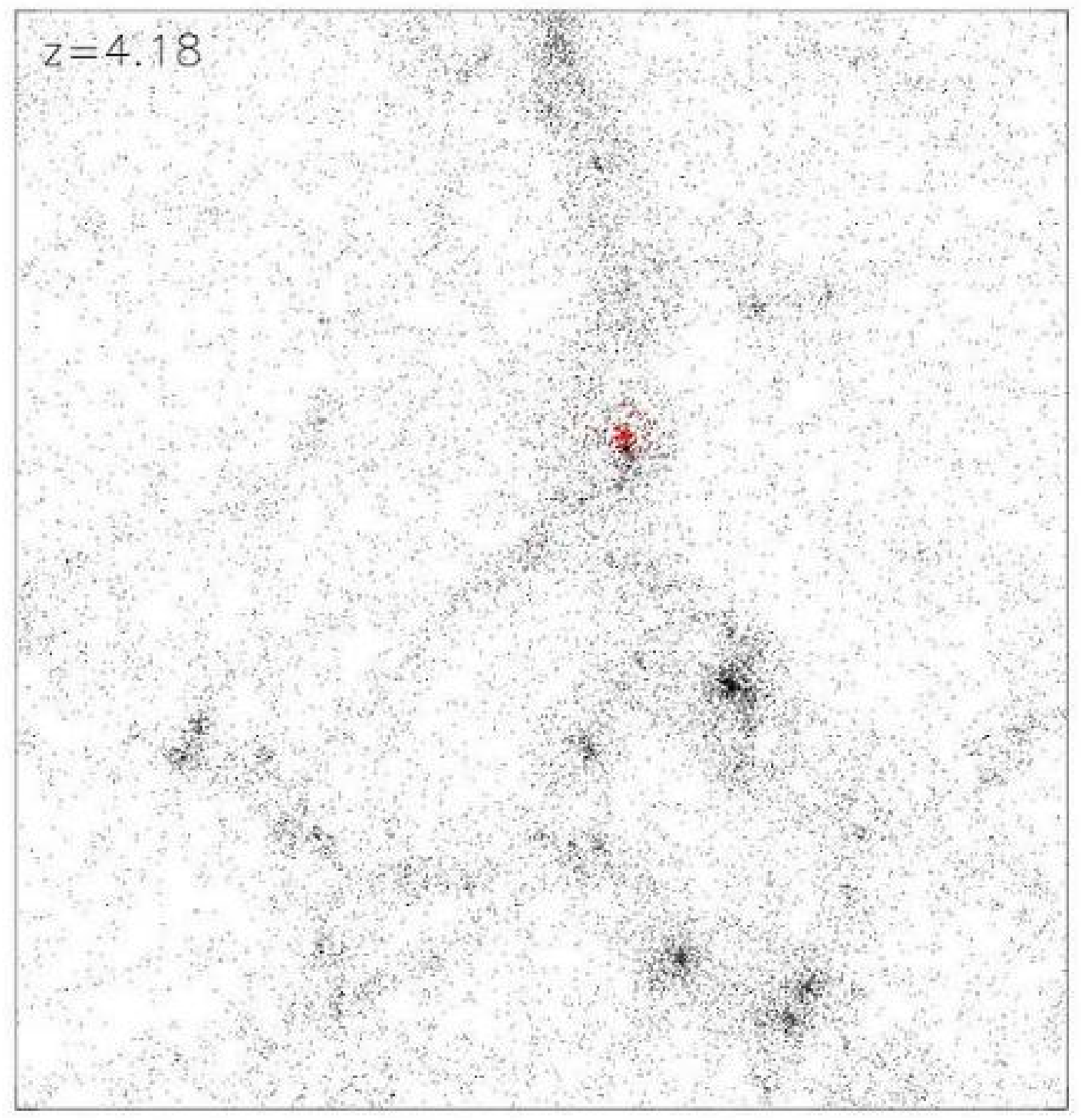}\plotone{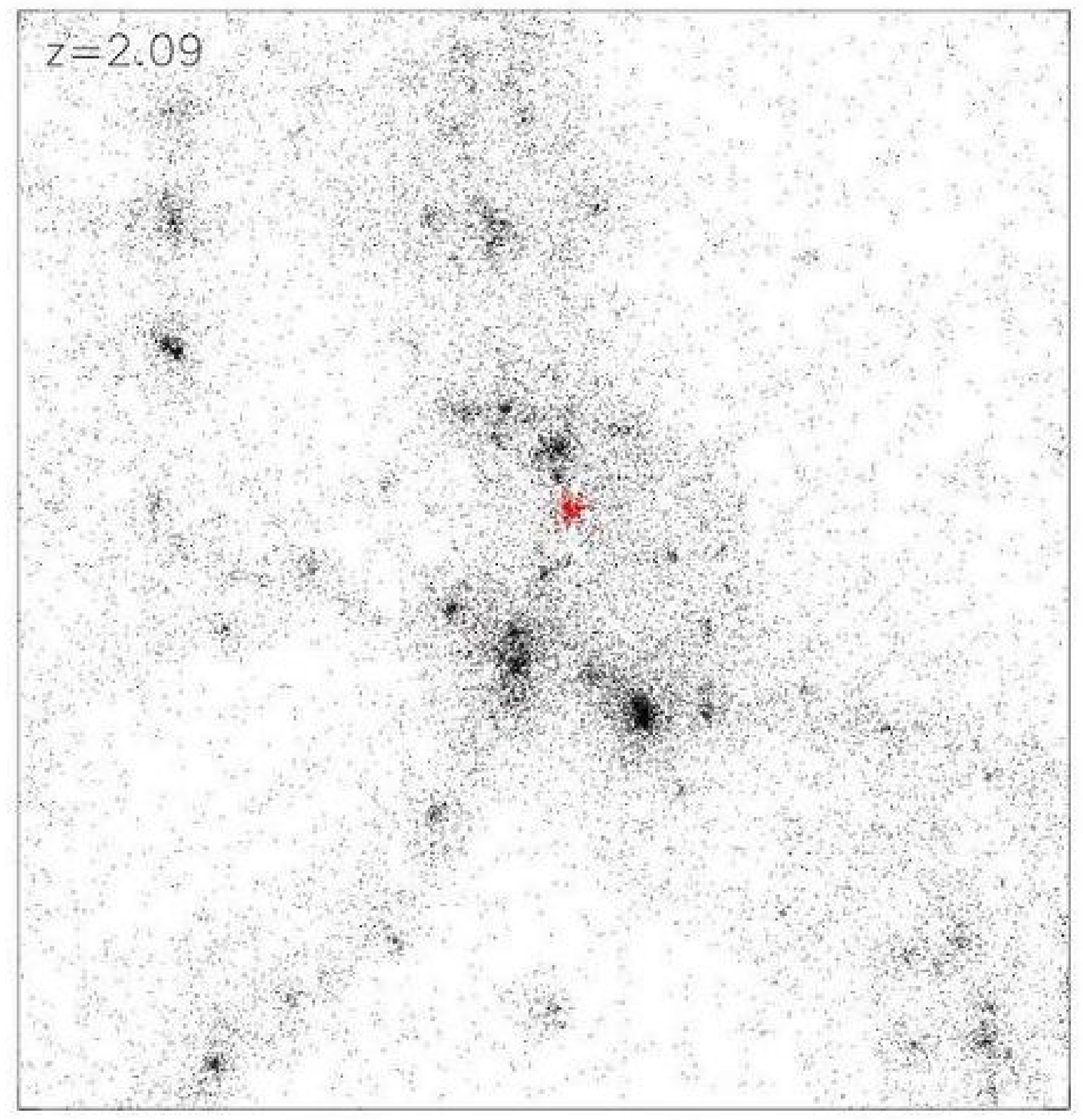}
\epsscale{0.4}\plotone{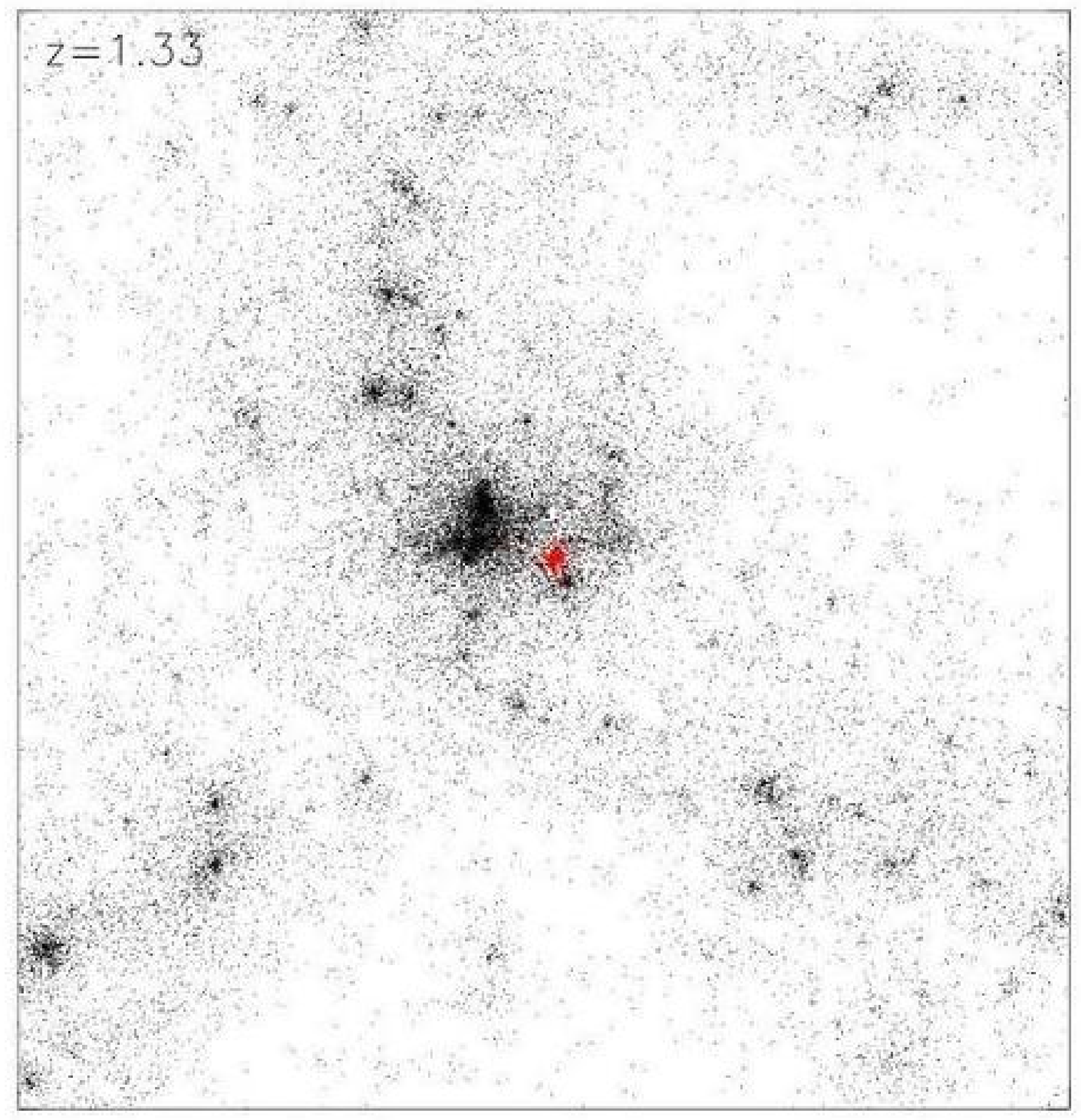}\plotone{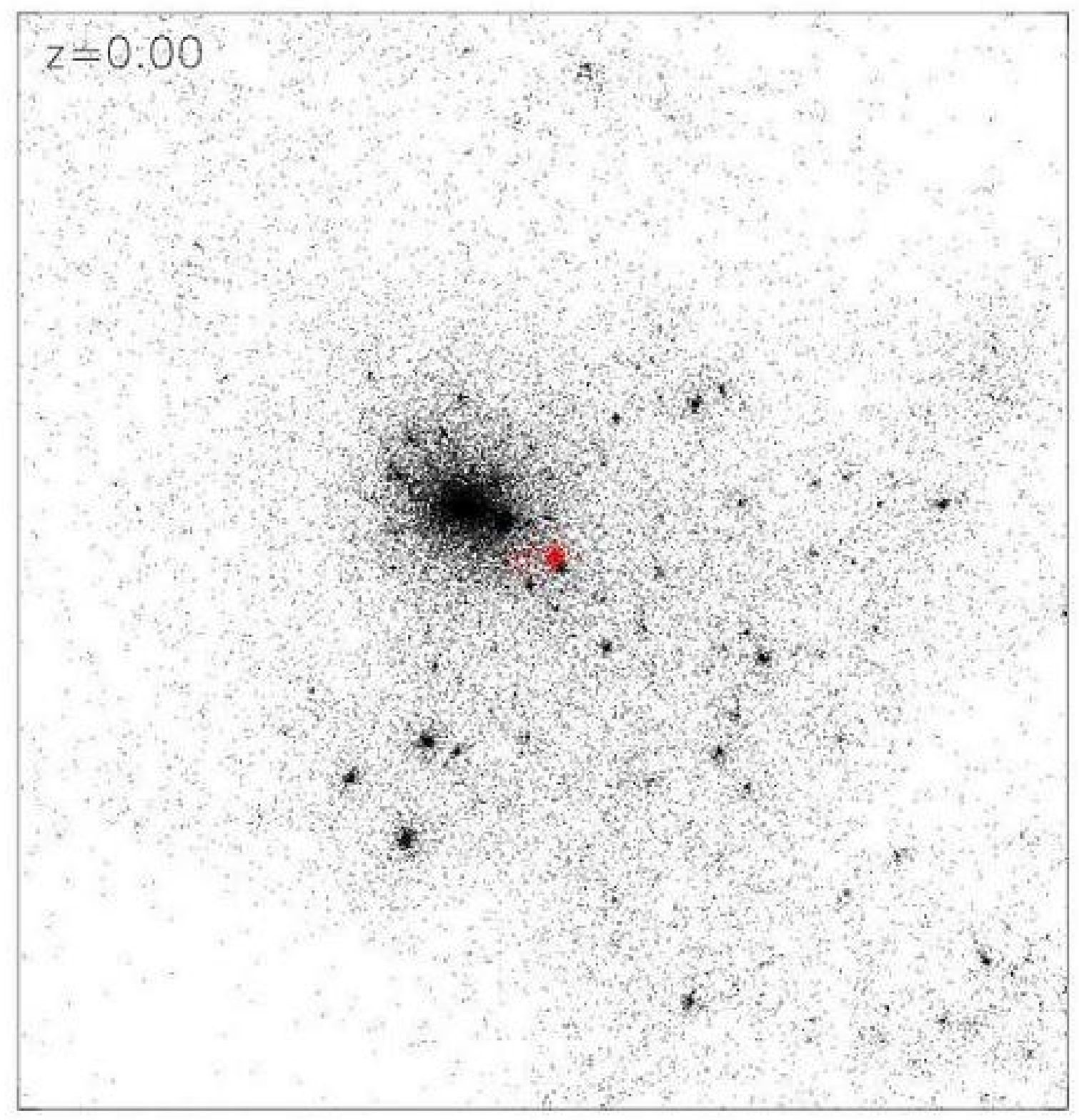} \caption{CDM
particle distributions around a low-mass old halo (with final mass
$M_h=1.2\times10^{11}h^{-1}{\rm M}_\odot$ and formation redshift
3.86) at a number of different redshifts (as indicated in the
panels). Particles that eventually end up in the final halo are
plotted in red.} \label{fig:5}
\end{figure*}
\begin{figure*}
\epsscale{0.4}\plotone{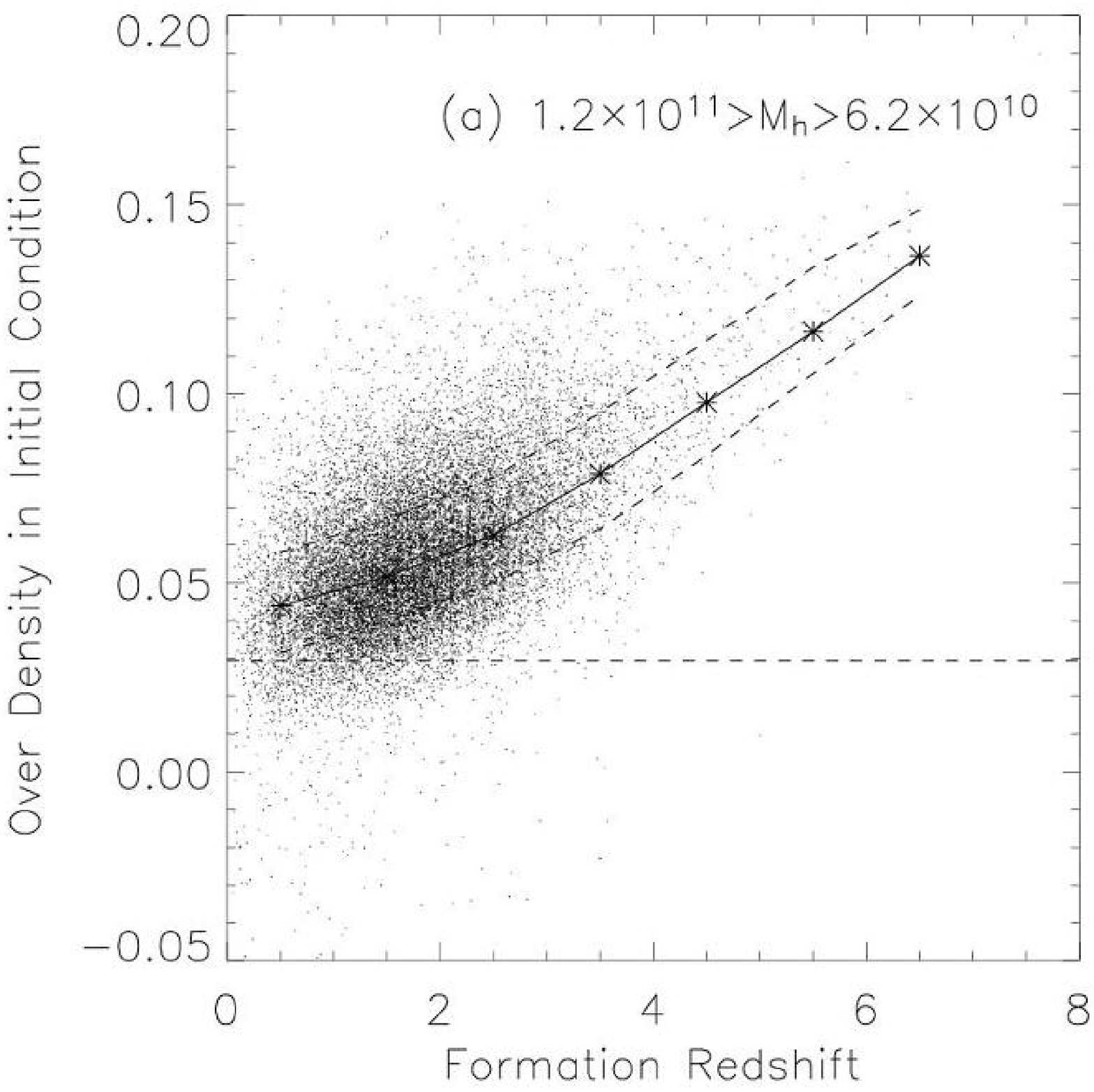}\plotone{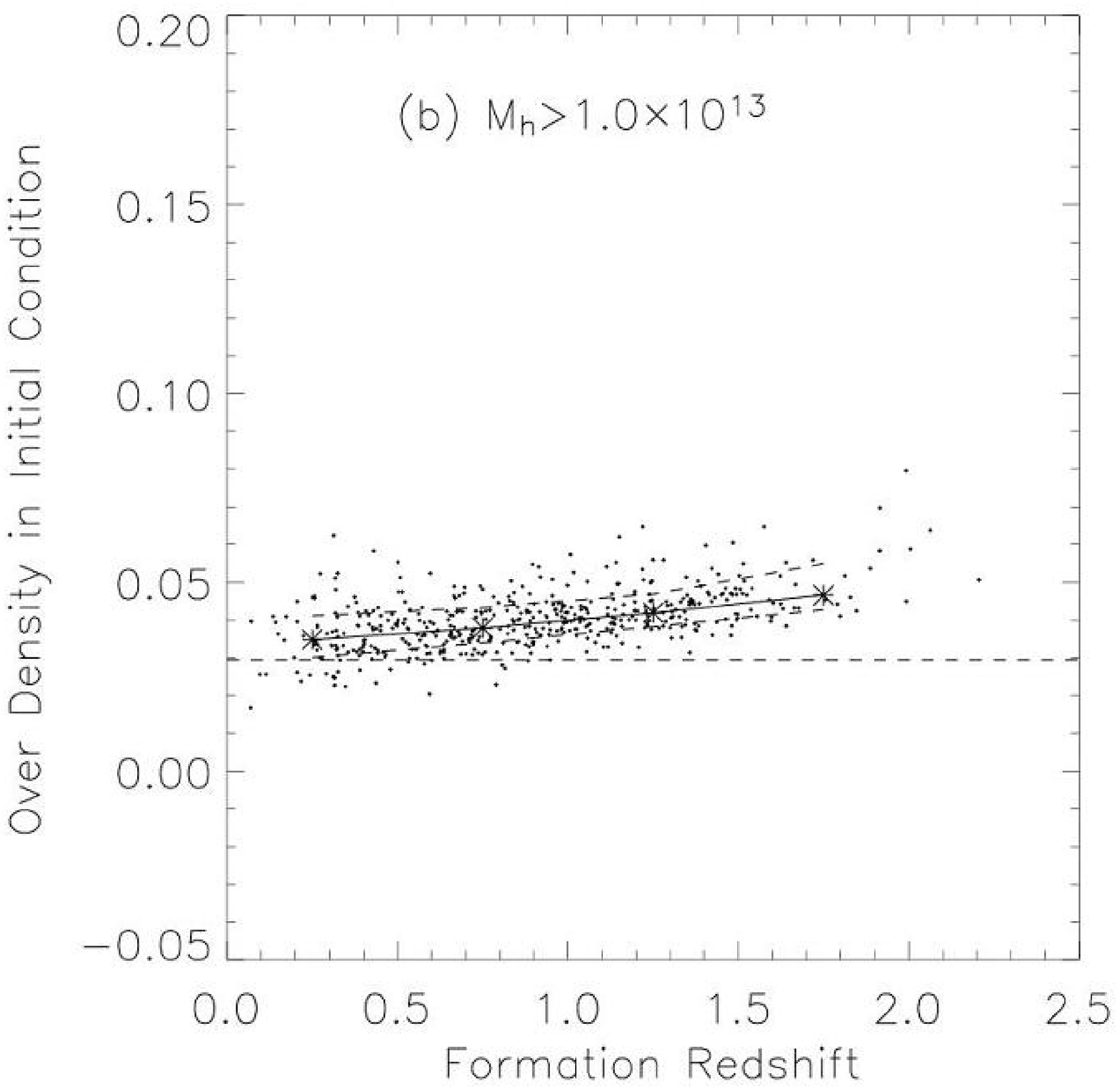} \caption{The
correlation between the initial overdensity, defined on the halo
mass scale, and halo formation redshift. Results are shown for
low-mass halos with $1.2\times 10^{11} h^{-1}{\rm
M}_\odot>M_h>6.2\times 10^{10} h^{-1}{\rm M}_\odot$ (left panel)
and for massive halos with $M_h> 1.0 \times 10^{13}h^{-1}{\rm
M}_\odot$ (right panel). The three curves in each panel show the
median, 20 and 80 percentiles. The horizontal lines indicates the
critical overdensity for collapse (at the initial redshift $z=72$)
based on spherical collapse model.} \label{fig:6}
\end{figure*}
\begin{figure*}
\epsscale{0.4}\plotone{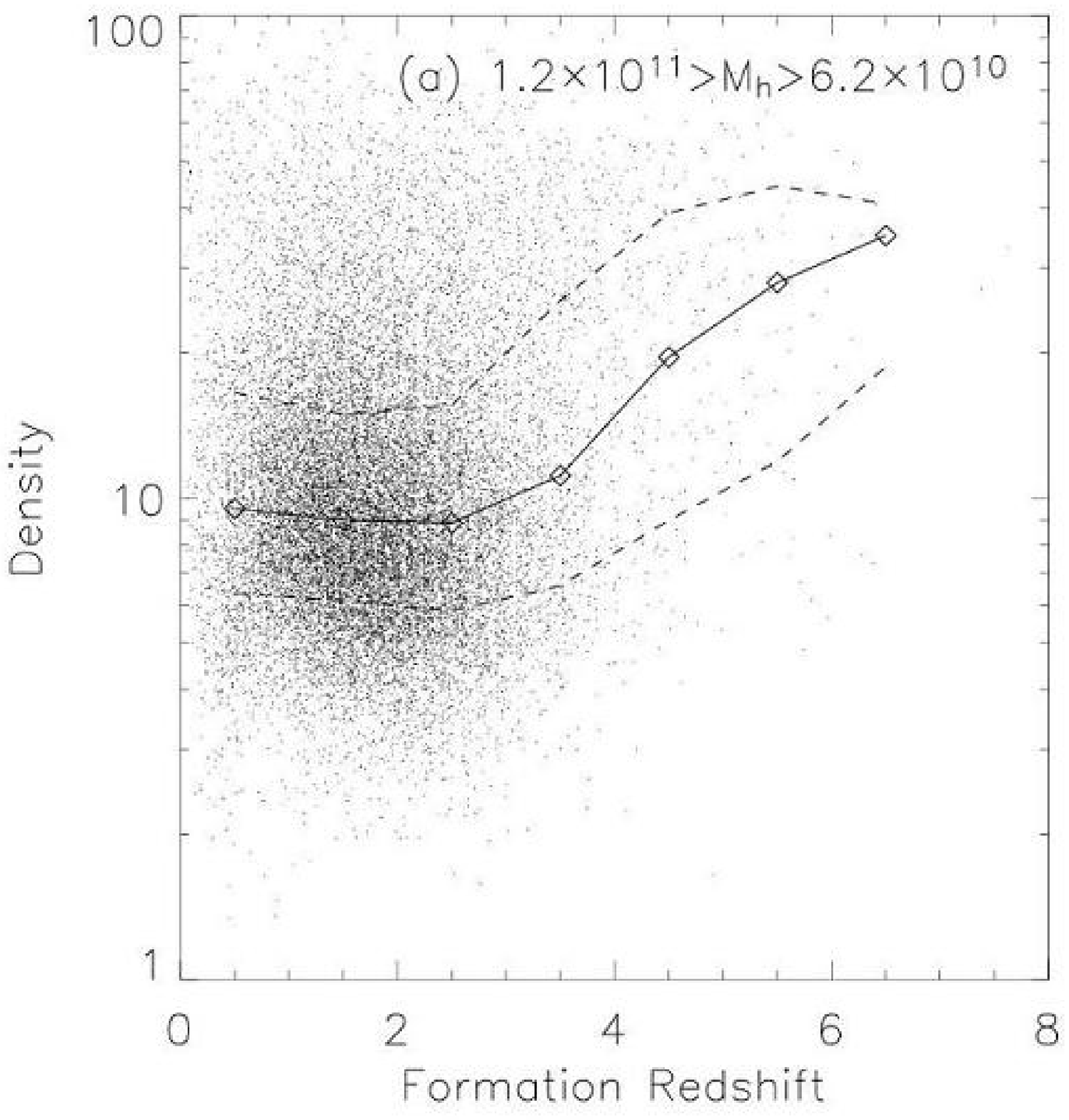} \plotone{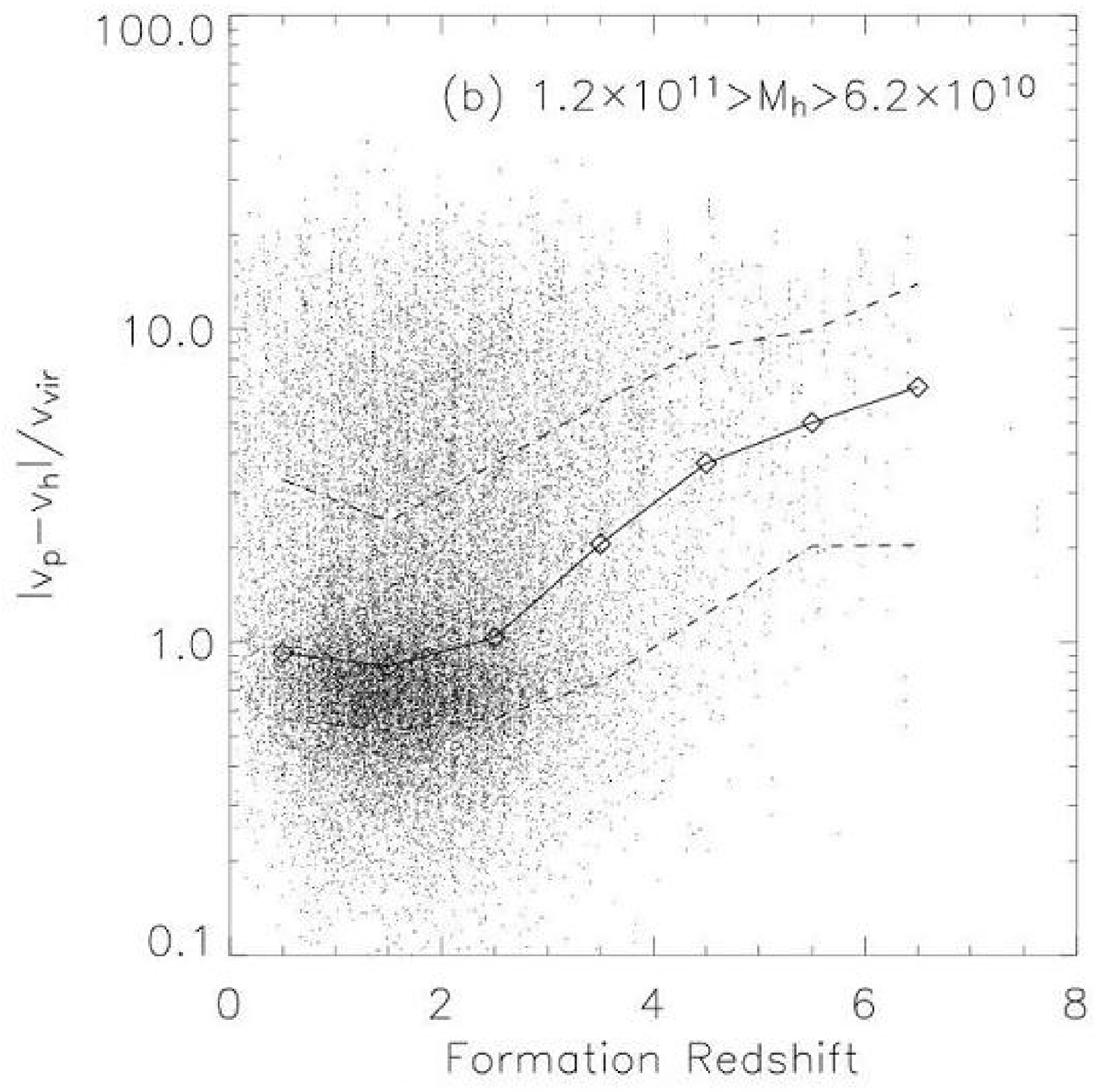}
\plotone{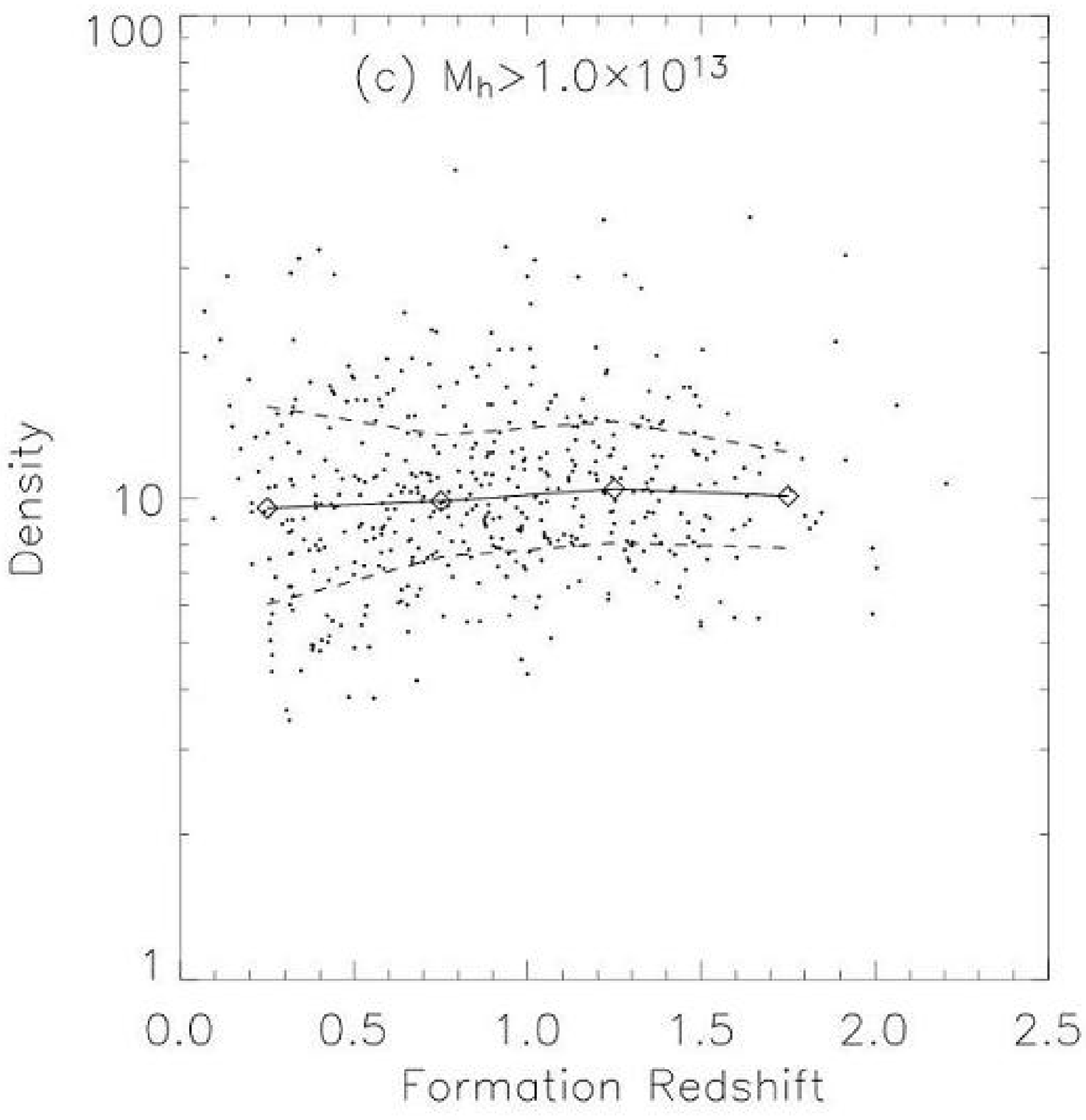}\plotone{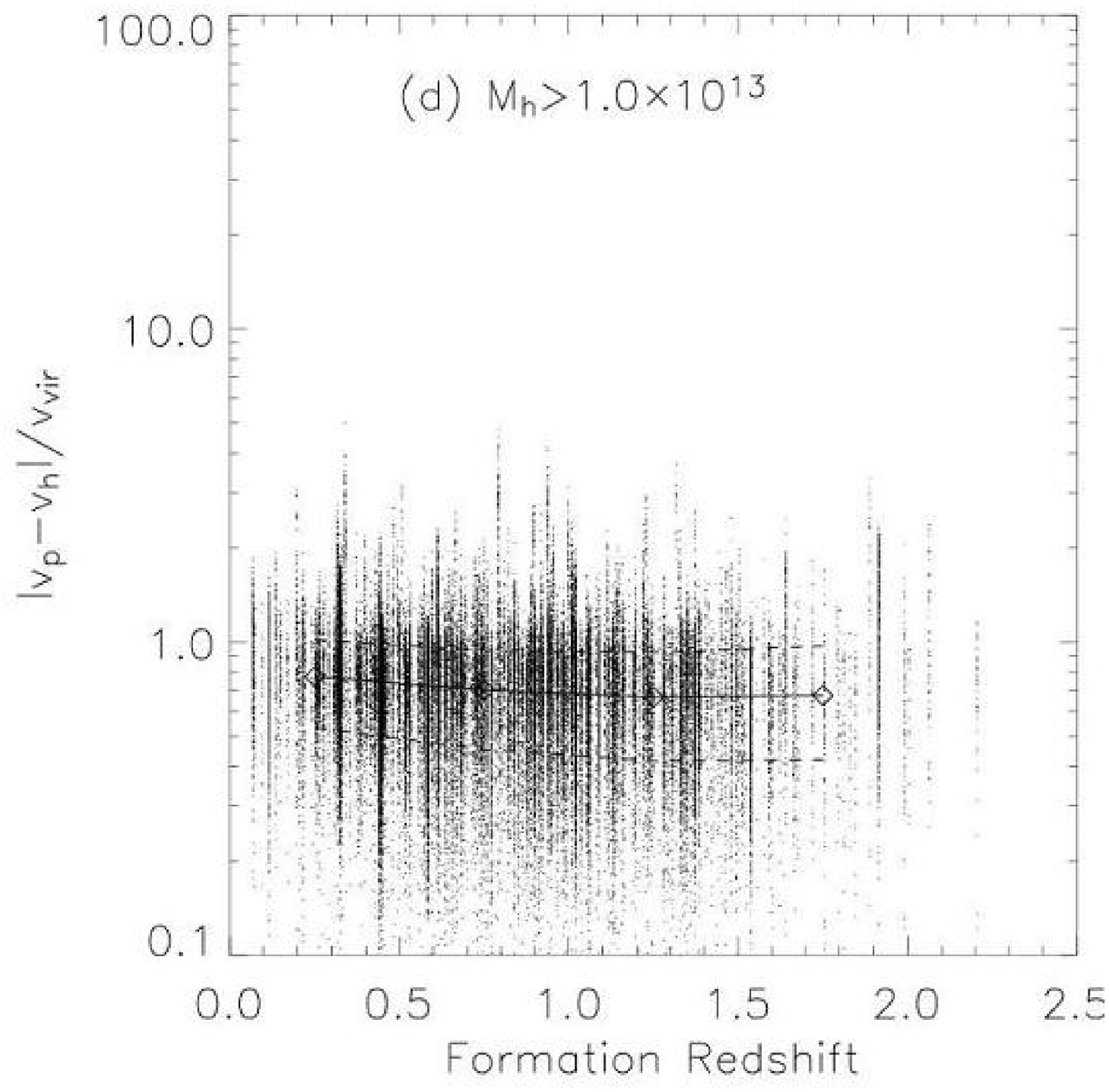} \caption{Left panels: the mean
density (in units of the mean density of the universe) between
$r_{vir}$ and $3r_{vir}$ as a function of halo formation redshift.
Each point represents a halo. Right panels: the velocity
difference for particles located between $r_{vir}$ and $3r_{vir}$
as a unction of halo formation redshift. Here each point
represents a dark matter particle. Upper panels are for low-mass
halos with $1.2\times 10^{11} h^{-1}{\rm M}_\odot>M_h>6.2\times
10^{10} h^{-1}{\rm M}_\odot$, and lower panels and for massive
halos with $M_h> 1.0 \times 10^{13}h^{-1}{\rm M}_\odot$. The three
curves in each panel are the median, 20 and 80 percentiles.}
\label{fig:7}
\end{figure*}
\begin{figure*}
\epsscale{1}\plotone{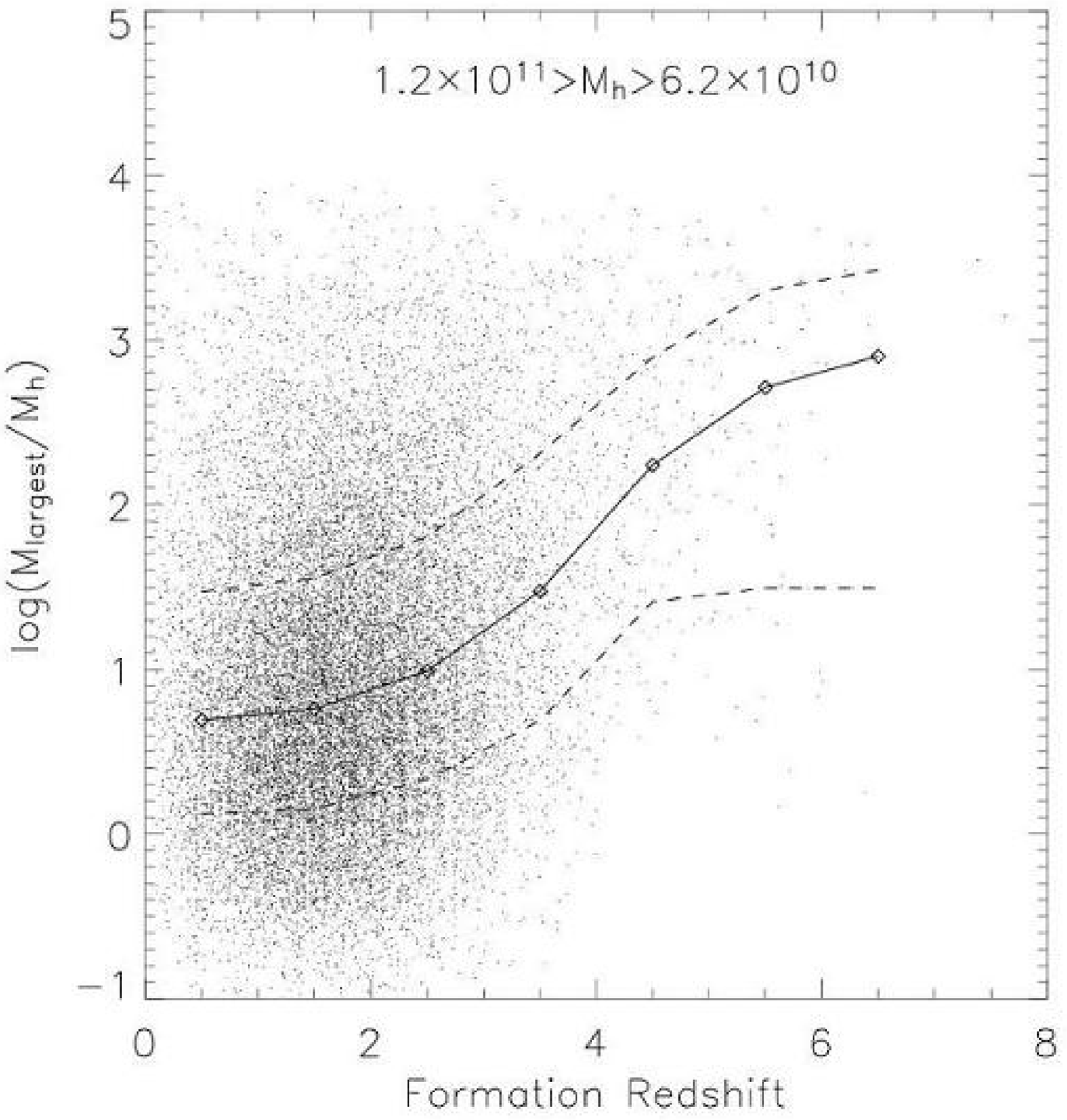} \caption{The ratio between $M_{\rm
largest}$ and $M_h$ as a function of halo formation redshift. Here
$M_{\rm largest}$ is defined to be the mass of the most massive
halo that is within $2h^{-1}{\rm Mpc}$ from the halo in
consideration. The three curves show the median, 20 and 80
percentiles. } \label{fig:8}
\end{figure*}

In order to understand the origin of the age-dependence of halo
clustering, we examine the properties of dark matter halos in the
initial density field. For each halo identified at the present
time ($z=0$), we trace the positions of all its particles back to
the initial condition, and define the initial center of the halo
to be the barycenter of these particles in the initial density
field (set at $z_{\rm i}=72$). At each center, we find the largest
spherical region within which the initial average overdensity
exceeds the critical value $\delta_{\rm sc}$ for spherical
collapse at $z=0$. We call this sphere the initial patch of the
halo, and denote the mass within it by $M_{\rm i1}$. For a given
halo, its initial center may be contained in the initial patches
of other halos, and we denote the mass within the largest patch by
$M_{\rm i}$. By definition $M_{\rm i1}\le M_{\rm i}$. According to
Press-Schechter theory (Press \& Schechter 1974), the mass of the
halo in which the center in consideration will eventually end up
should be equal to $M_{\rm i}$ instead of $M_{\rm i1}$. The reason
for this is discussed in detail in Sheth, Mo \& Tormen (2001). In
Figs.\,\ref{fig:2} we show the ratio between the initial mass
$M_{\rm i}$ and the final halo mass, $M_h$, versus halo formation
time for halos in two mass ranges. These plots show that, for many
low-mass halos, the initial mass assigned according to the
spherical collapse model is much larger than the final halo mass.
Note that the initial/final mass ratio can be as large as 1,000
for some low-mass halos. This discrepancy between the
Press-Schechter theory and cosmological simulations has been
discussed in considerable detail in Sheth et al. (2001), and is
the primary reason why Press-Schechter formula over-predicts the
abundance of low-mass halos. For high-mass halos, the ratios are
much closer to 1, suggesting that Press-Schechter formalism works
better for high-mass halos.

  Fig.\,\ref{fig:2} also shows a clear trend that the
initial/final mass ratio on average increases with the formation
redshift, and the trend is stronger for halos of lower masses.
According to the halo bias model of Mo \& White (1996), which is
based on identifying halos in the initial density field according
to spherical collapse model, halos of larger masses should be more
strongly correlated. It is therefore interesting to check whether
the age-dependence of halo clustering observed in the simulation
can be explained by the correlation between initial mass and
formation time obtained here. To do this, we select low-mass halos
with final masses $1.2\times 10^{11} h^{-1}{\rm
M}_\odot>M_h>6.2\times 10^{10} h^{-1}{\rm M}_\odot$, and massive
halos with final masses $M_h> 10^{13}h^{-1}{\rm M}_\odot$. We bin
halos in each of the two mass ranges according to their initial
masses $M_{\rm i}$, and compute the bias factor as a function of
$M_{\rm i}$. The results are shown in Fig.\,\ref{fig:3}. As
expected, the bias factor increases strongly with mass. The
prediction of the Mo \& White (1996) bias model is shown as the
solid curve. The model matches the simulation result for $M_{\rm
i}\gg M_\star$, suggesting that large spherical patches are
strongly clustered regardless of whether or not a small halo will
eventually form in them. However, for $M_{\rm i}\la M_\star$, the
model overpredicts the bias factor. This is not surprising,
because small patches in which a low-mass halo form at the present
time are biased towards low-density regions.

  We can assign to each halo a bias factor, $b$, according to
its initial mass $M_{\rm i}$ using the bias-mass relation given by
the simulation. We can then average the values of $b$ for halos in
bins of formation redshift to predict the age-dependence of halo
clustering. The dashed curves in Fig.\,\ref{fig:4} show the
bias-formation redshift relations obtained in this way for halos
in two mass ranges. The predictions match well the simulation
results, suggesting that the observed age-dependence of clustering
for halos of a given final mass can be entirely explained by the
difference in the initial mass. Old, low-mass halos are more
strongly clustered than their younger counterparts simply because
they are associated with perturbations of higher mass (i.e.
patches corresponding to higher peaks) in the initial density
field. The age-dependence is weaker for more massive halos,
because their $M_{\rm i}$ represent their final masses more
faithfully.

\section{The environmental effects of halo formation}
\label{sec:env}

The question is, of course, why some low-mass halos can survive in
a patch which itself should collapse to form a virialized halo
according to spherical collapse model, and what prevents such
halos from accreting mass in later times. To have some ideas about
what is going on during the formation of a low-mass old halo, we
show in Fig.\,\ref{fig:5} the spatial distribution of dark matter
particles in the neighborhood of such a halo for a number of
representative snapshots, with the particles that eventually end
up in the final halo marked in colour. As one can see, the main
part of the small halo collapses quickly to form a halo, which
then grows only slowly while orbiting around a bigger halo that
forms later. The early collapse is driven largely by the
relatively high initial density of the material associated with
the halo, as is demonstrated in Fig.\,\ref{fig:6} which shows the
initial overdensity around a halo center as a function of halo
formation redshift. In Fig.\,\ref{fig:6} the overdensity is
defined as the mean within a sphere that is centered at the
initial position of the halo and has a radius such that the total
mass it contains is equal to the mass of the halo. For low-mass
halos, there is a strong, positive correlation between the mean
overdensity and the formation time. Halos with earlier formation
times have initial overdensity much higher than the critical
density for spherical collapse (indicated as the horizontal line
in the figure). With such overdensities, these perturbations can
collapse in early times to form virialized halos. The situation is
very different for massive halos, where the mean overdensity are
all close to the critical value for spherical collapse [see
Fig.\,\ref{fig:6}(b)].

  There are two possibilities why low-mass old halos stop accreting
mass at relatively early times. The first is that these halos
reside in locally low-density regions, and so there is no material
for them to accrete. To test this, we estimate the total mass
between $r_{\rm vir}$ and $3 r_{\rm vir}$ (with $r_{\rm vir}$ the
virial radius) for each halo and plot the mean mass density (in
units of mean density of the universe) within these two radii
versus halo formation redshift in Fig.\,\ref{fig:7}(a). This
figure shows that there is in fact a slightly larger amount of
mass around older halos,  and so the truncation of the growth of
old, low-mass halos cannot be explained by the lack of material
around them. However, the availability of material is a necessary,
not a sufficient condition for halo accretion. In order for the
material around a halo to be accreted by the halo, the velocity of
the material to be accreted relative to the halo must be
sufficiently low. To see if this condition is fulfilled, we
estimate the velocity difference $\vert {\bf v}_{p}-{\bf v}_{\rm
h}\vert$ for all particles with halo-centric distances between
$r_{\rm vir}$ and $3 r_{\rm vir}$, where ${\bf v}_{\rm h}$ is the
velocity of the host halo, and ${\bf v}_{p}$ is the velocity of a
CDM particle. Fig.\,\ref{fig:7}(b) shows the velocity difference
(in units of halo virial velocity) versus halo formation redshift.
There is a clear trend that particles in the neighborhood of an
older halo have systematically larger velocity differences with
the host halo. For the oldest halos, the velocity differences for
most particles are significantly larger than the virial velocities
of the halos. These particles cannot be accreted by the halo,
because they are too energetic. Thus, the reason why late
accretions by old, low-mass halos are truncated is that they are
embedded in `hot' regions where particles are moving too fast for
them to capture. The situation is very different for massive halos
[Figs.\,\ref{fig:7}(c) and (d)]. First of all, the mean densities
in their neighborhoods are quite independent of halo formation
redshift [Fig.\,\ref{fig:7}(c)]. More importantly, the velocity
differences between a halo and particles around it are smaller
than the virial velocity and so most of these particles can be
accreted by the halo in subsequent evolution.

  It is easy to understand why old, low-mass halos are embedded in
hot environments. As shown in Fig.\,\ref{fig:5}, these halos are
formed in the vicinity of big structures that are dominated by one
or more massive halos. To see this more clearly, we plot in
Fig.\,\ref{fig:8} the mass of the most massive neighboring halo
within a distance of $2 h^{-1}{\rm Mpc}$ from a low-mass halo
as a function of the formation redshift of the low-mass halo.
An old low-mass halo almost always has a massive halo in its
neighborhood. These old low-mass halos are
similar to the subhalo population see in high-resolution $N$-body
simulations (e.g. Klypin et al. 1999; Moore et al. 1999), except that
they have not yet merged into the virialized part of a massive halo.
Although the formation of the more massive neighboring halos occurred
at later times, the tidal field of the large structure can accelerate
the particles around low-mass halos, thereby increasing their velocities
relative to the low-mass halos. Furthermore, massive pancakes and
filaments can form in high-density patches prior to the formation of massive
halos, and heat up the particles in them in a way similar to the
preheating mechanism discussed in Mo et al. (2005). Note that
these two processes are related, because the formation of pancakes
and filaments is related to the local tidal field. As discussed in
Mo et al. (2005), the tidal effect and preheating are expected to
have more significant impact on smaller halos, which may explain
why the age-dependence of halo clustering is weaker for more
massive halos.

\section {Discussion and summary}
\label{sec:discussion}

In this paper, we have used a high-resolution $N$-body simulation
to study how the formation of CDM halos may be affected by their
environments, and how such environmental effects may produce the
age-dependence of halo clustering seen in recent large $N$-body
simulations. We have shown that old low-mass halos are almost
always associated with initial perturbations that correspond to
higher peaks than their present masses imply. Consequently, these
halos are strongly clustered in space despite of their low masses
at the present time. Because of their associations with large
density perturbations, old low-mass halos are almost always found
in the vicinity of big structures that are dominated by one or
more massive halos at the present time. Mass accretions into these
low-mass halos are limited at late times by the tidal field of the
larger neighboring structures. Such environmental effects are
weaker for more massive halos, which explains why the
age-dependence of halo clustering is weaker for more massive
halos.

   Our results demonstrate clearly that environmental effects play an
important role during the formation of dark matter halos,
especially of low-mass halos. Since the properties of galaxies
that form in a halo are expected to depend on the formation
history of the halo, such environmental effects may have important
implications for galaxy formation in different environments.
Galaxies that form in old, low-mass halos are presumably old and
faint, and our results suggest that many such galaxies may be
located in the vicinity of relatively massive galaxy systems.
These galaxies likely  represent an extension of the satellite
galaxies observed in galaxy systems such as the local group and
other groups/clusters of galaxies. Observationally, it is known
that satellite galaxies in galaxy groups are dynamically old
systems (dwarf ellipticals and dwarf spheroids) that contain
mainly old stellar populations. It is interesting to see whether
the stellar populations and kinematics of low-mass galaxies in the
neighborhoods of nearby galaxy groups have properties similar to
those of the satellite population.

 Theoretically, it is very interesting to see if the ellipsoidal
collapse model that incorporates large-scale tidal field into the
dynamics can indeed explain the environmental effects discussed
here and the age-dependence of halo clustering observed in
$N$-body simulations. We will come back to this problem in a
future work.


\section*{Acknowledgment}
We thank Xi Kang and Guangtun Zhu for help in constructing merger
trees from the simulation, Adrian Jenkins, the referee of the paper,
for helpful comments.
HJM would like to acknowledge the support
of NSF AST-0607535, NASA AISR-126270 and NSF IIS-0611948.  YPJ is
supported by the grants from NSFC (No. 10373012, 10533030) and from
Shanghai Key Projects in Basic research (No. 04JC14079 and
05XD14019). HYW and HJM acknowledge supports from the Chinese Academy
of Sciences for the visit at Shanghai Astronomical Observatory.


\label{lastpage}


\begin{thebibliography}{}

\bibitem[\protect\citeauthoryear{Bardeen et
al.}{1986}]{1986ApJ...304...15B} Bardeen J.~M., Bond J.~R., Kaiser
N., Szalay A.~S., 1986, ApJ, 304, 15

\bibitem[\protect\citeauthoryear{Bond et al.}{1991}]{1991ApJ...379..440B}
Bond J.~R., Cole S., Efstathiou G., Kaiser N., 1991, ApJ, 379, 440

\bibitem[\protect\citeauthoryear{Davis et al.}{1985}]{1985ApJ...292..371D}
Davis M., Efstathiou G., Frenk C.~S., White S.~D.~M., 1985, ApJ,
292, 371

\bibitem[\protect\citeauthoryear{Gao, Springel, \&
White}{2005}]{2005MNRAS.363L..66G} Gao L., Springel V., White
S.~D.~M., 2005, MNRAS, 363, L66

\bibitem[\protect\citeauthoryear{Harker et al.}{2006}]{2006MNRAS.367.1039H}
Harker G., Cole S., Helly J., Frenk C., Jenkins A., 2006, MNRAS,
367, 1039

\bibitem[\protect\citeauthoryear{Jing}{1998}]{1998ApJ...503L...9J} Jing
Y.~P., 1998, ApJ, 503, L9

\bibitem[Jing(2000)]{jing00} Jing, Y.~P.\ 2000, \apj, 535, 30

\bibitem[]{JMB98}
Jing Y.P., Mo H.J., B\"{o}rner G., 1998, \apj , 494, 1

\bibitem[\protect\citeauthoryear{Jing \& Suto}{2002}]{2002ApJ...574..538J}
Jing Y.~P., Suto Y., 2002, ApJ, 574, 538

\bibitem[]{JS00a}
Jing Y.P., Suto Y., 2000, \apj , 529L, 69

\bibitem[]{Klypin99}
Klypin A. Gottl{\"o}ber S., Kravtsov A.V., Khokhlov A.M.,
1999, \apj , 516, 530

\bibitem[\protect\citeauthoryear{Li, Mo, \& van den
Bosch}{2005}]{2005astro.ph.10372L} Li Y., Mo H.~J., van den Bosch
F.~C., 2005, preprint, arXiv:astro-ph/0510372

\bibitem[\protect\citeauthoryear{Lu et al.}{2006}]{2006MNRAS.368.1931L} Lu
Y., Mo H.~J., Katz N., Weinberg M.~D., 2006, MNRAS, 368, 1931

\bibitem[\protect\citeauthoryear{Ma \& Fry}{2000}]{2000ApJ...543..503M} Ma
C.-P., Fry J.~N., 2000, ApJ, 543, 503

\bibitem[\protect\citeauthoryear{Mo, Jing, \&
B\"orner}{1997}]{1997MNRAS.286..979M} Mo H.~J., Jing Y.~P., Borner
G., 1997, MNRAS, 286, 979

\bibitem[\protect\citeauthoryear{Mo \& White}{1996}]{1996MNRAS.282..347M}
Mo H.~J., White S.~D.~M., 1996, MNRAS, 282, 347

\bibitem[\protect\citeauthoryear{Mo et al.}{2005}]{2005MNRAS.363.1155M} Mo
H.~J., Yang X., van den Bosch F.~C., Katz N., 2005, MNRAS, 363,
1155

\bibitem[]{Moore99}
Moore B., Ghigna S., Governato G., Lake G., Quinn T., Stadel J.,
Tozzi P., 1999, ApJ., 524, L19

\bibitem[\protect\citeauthoryear{Navarro, Frenk, \&
White}{1997}]{1997ApJ...490..493N} Navarro J.~F., Frenk C.~S.,
White S.~D.~M., 1997, ApJ, 490, 493

\bibitem[\protect\citeauthoryear{Peacock \&
Smith}{2000}]{2000MNRAS.318.1144P} Peacock J.~A., Smith R.~E.,
2000, MNRAS, 318, 1144

\bibitem[\protect\citeauthoryear{Press \&
Schechter}{1974}]{1974ApJ...187..425P} Press W.~H., Schechter P.,
1974, ApJ, 187, 425

\bibitem[\protect\citeauthoryear{Seljak}{2000}]{2000MNRAS.318..203S} Seljak
U., 2000, MNRAS, 318, 203

\bibitem[\protect\citeauthoryear{Seljak \&
Warren}{2004}]{2004MNRAS.355..129S} Seljak U., Warren M.~S., 2004,
MNRAS, 355, 129

\bibitem[\protect\citeauthoryear{Sheth, Mo, \&
Tormen}{2001}]{2001MNRAS.323....1S} Sheth R.~K., Mo H.~J., Tormen
G., 2001, MNRAS, 323, 1

\bibitem[\protect\citeauthoryear{Wechsler et
al.}{2002}]{2002ApJ...568...52W} Wechsler R.~H., Bullock J.~S.,
Primack J.~R., Kravtsov A.~V., Dekel A., 2002, ApJ, 568, 52

\bibitem[\protect\citeauthoryear{Wechsler et
al.}{2005}]{2005astro.ph.12416W} Wechsler R.~H., Zentner A.~R.,
Bullock J.~S., Kravtsov A.~V., Allgood B., 2005, preprint,
arXiv:astro-ph/0512416

\bibitem[\protect\citeauthoryear{Wetzel et al.}{2006}]{2006astro.ph..6699W}
Wetzel A.~R., Cohn J.~D., White M., Holz D.~E., Warren M.~S.,
2006, preprint, arXiv:astro-ph/0606699

\bibitem[\protect\citeauthoryear{Yang, Mo, \& van den
Bosch}{2003}]{2003MNRAS.339.1057Y} Yang X., Mo H.~J., van den
Bosch F.~C., 2003, MNRAS, 339, 1057

\bibitem[\protect\citeauthoryear{Zhao et al.}{2003}]{2003ApJ...597L...9Z}
Zhao D.~H., Jing Y.~P., Mo H.~J., B{\"o}rner G., 2003, ApJ, 597,
L9

\bibitem[\protect\citeauthoryear{Zhao et al.}{2003}]{2003MNRAS.339...12Z}
Zhao D.~H., Mo H.~J., Jing Y.~P., B{\"o}rner G., 2003, MNRAS, 339,
12

\bibitem[\protect\citeauthoryear{Zheng et al.}{2005}]{2005ApJ...633..791Z}
Zheng Z., et al., 2005, ApJ, 633, 791

\bibitem[\protect\citeauthoryear{Zhu et al.}{2006}]{2006ApJ...639L...5Z}
Zhu G., Zheng Z., Lin W.~P., Jing Y.~P., Kang X., Gao L., 2006,
ApJ, 639, L5

\end{thebibliography}
\end{document}